# Three-color Sagnac source of polarization-entangled photon pairs

**Michael Hentschel,**[1,3,*] **Hannes Hübel,**[2] **Andreas Poppe**[2,3]**, and Anton Zeilinger**[1,2]

[1]*Institute for Quantum Optics and Quantum Information - IQOQI, Austrian Academy of Sciences, Boltzmanngasse 3, 1090 Vienna, Austria*
[2]*Quantum Optics, Quantum Nanophysics and Quantum Information, Faculty of Physics, University of Vienna, Boltzmanngasse 5, 1090 Vienna, Austria*
[3]*Austrian Institute of Technology GmbH - AIT, Safety & Security Department, Donau-City-Str. 1, 1220 Vienna, Austria*
[*]*michael.hentschel@ait.ac.at*

**Abstract:** We demonstrate a compact and stable source of polarization-entangled pairs of photons, one at 810 nm wavelength for high detection efficiency and the other at 1550 nm for long-distance fiber communication networks. Due to a novel Sagnac-based design of the interferometer no active stabilization is needed. Using only one 30 mm ppKTP bulk crystal the source produces photons with a spectral brightness of $1.13 \times 10^6$ pairs/s/mW/THz with an entanglement fidelity of 98.2%. Both photons are single-mode fiber coupled and ready to be used in quantum key distribution (QKD) or transmission of photonic quantum states over large distances.

## References and links

1. P. Kwiat, K. Mattle, H. Weinfurter, A. Zeilinger, A. Sergienko, and Y. Shih, "New high-intensity source of polarization-entangled photon pairs," Phys. Rev. Lett. **75**, 4337-4341 (1995).
2. E. Mason, M. Albota, F. König, and F. Wong, "Efficient generation of tunable photon pairs at 0.8 and 1.6 μm," Opt. Lett. 27, 2115-2117 (2002).
3. S. M. Spillane, M. Fiorentino, and R. Beausoleil, "Spontaneous parametric down conversion in a nanophotonic waveguide," Opt. Express **15**, 8770-8780 (2007).
4. S. Zhang, J. Yao, W. Liu, Z. Huang, J. Wang, Y. Li, C. Tu, and F. Lu, "Second harmonic generation of periodically poled potassium titanyl phosphate waveguide using femtosecond laser pulses," Opt. Express **16**, 14180-14185 (2008).
5. G. Ribordy, J. Brendel, J. Gautier, N. Gisin, and H. Zbinden, "Long-distance entanglement-based quantum key distribution," Phys. Rev. A **63**, 012309 (2000).
6. A. Treiber, A. Poppe, M. Hentschel, D. Ferrini, T. Lorünser, E. Querasser, T. Matyus, H. Hübel, and A. Zeilinger, "Fully automated entanglement-based quantum cryptography system for telecom fiber networks," New J. Phys. **11**, 045013 (2009).
7. D. Ljunggren, and M. Tengner, "Optimal focusing for maximal collection of entangled narrow-band photon pairs into single-mode fibers," Phys. Rev. A **72**, 062301 (2005).
8. H. Hübel, M. Vanner, T. Lederer, B. Blauensteiner, T. Lorünser, A. Poppe, and A. Zeilinger, "High-fidelity transmission of polarization encoded qubits from an entangled source over 100 km of fiber," Opt. Express **15**, 7853-7862 (2007).
9. A. Fedrizzi, T. Herbst, A. Poppe, T. Jenewein, and A. Zeilinger, "A wavelength-tunable fiber-coupled source of narrowband entangled photons," Opt. Express **15**, 15377-15386 (2007).
10. S. Sauge, M. Swillo, M. Tengner, and A. Karlsson, "A single-crystal source of path-polarization entangled photons at non-degenerate wavelengths," Opt. Express **16**, 9701-9707 (2008).
11. Schott optical glass catalogue.
12. D. James, P. Kwiat, W. Munro, and A. White, "Measurement of qubits," Phys. Rev. A **64**, 052312 (2001).

## 1. Introduction

Entanglement is a phenomenon based on the very principles of quantum mechanics, leading to surprising consequences once described as "spooky action at a distance" by Albert Einstein. Techniques employing this phenomenon, particularly two-photon entanglement, have matured in the past years to become powerful and reliable tools in quantum optics and quantum information experiments such as quantum computing, quantum teleportation and QKD. As the most popular method to achieve entangled photon states spontaneous parametric down conversion (SPDC) in nonlinear media has progressed considerably in recent times. Originally, non-collinear emission from type-II bulk-BBO crystals was employed, where photons from two intersecting cones were collected to form the polarization-entangled two-photon state [1]. Due to the relatively small spatial overlap of the photon modes in this geometry such schemes suffered from a rather poor efficiency. When manufacturing technologies in nonlinear optics advanced high quality periodically poled crystals became available. Collinear schemes using quasi-phase matching were developed providing better efficiency and ease of alignment [2]. Here the full capabilities of the nonlinear material can be exploited by appropriate engineering of the crystal orientation and poling period. Especially the availability of bulk crystals exceeding several millimeters in length strongly accelerated the development of high-brightness sources. Recently, even waveguides have been written into periodically poled crystals for further enhancement of conversion efficiencies [3,4], which are however beyond the scope of this work.

Our choice of asymmetric SPDC with the signal wavelength close to the visible and the idler in the telecom region comes from the particular demands of real-world applications. In the case of QKD, the system should operate with optical fibers and also deliver a high key rate. Clearly, designs with both photons below 1 μm, like standard BBO sources, are not compatible with the telecom infrastructure, and sources where both signal and idler photons are at telecom wavelengths will not achieve high rates. The latter comes from the low detection efficiencies (~15%) and high noise figures of InGaAs-avalanche photo diodes (APDs) employed in the infrared. Single photons around 800 nm however can be detected with efficiencies of up to 50% with far lower background noise using Si-APDs. Hence asymmetric SPDC sources, first demonstrated in 2000 [5], combine the high detection efficiency in one arm (locally) with the low loss transmission in optical fibers for the other arm. As long as the detection problem in the infrared is not solved, asymmetric setups will be needed for applications requiring high coincidence rates *and* telecom compatibility.

A recent QKD experiment [6], based on a polarization-entangled asymmetric SPDC source, showed that such systems suffer from chromatic dispersion in the optical fiber. Since in that experiment the idler bandwidth was filtered to only around 3.6 nm, chromatic dispersion broadened the coincidence peak after 25 km of standard fiber to ~1.6 ns, which is comparable to the typical gate width of an InGaAs-APD. For longer fiber distances either counts will be lost or a larger gate width will lead to higher errors. Asymmetric SPDC source have therefore to be designed to yield a narrow bandwidth minimizing dispersion effects.

## 2. Design considerations

The photon flux and bandwidth generated and collected by the optical fibers inherently depend on design parameters such as crystal length, geometry of pump focusing, and fiber coupling arrangement. For collinear down conversion the photon emission occurs in a manifold of concentric cones, fulfilling momentum and energy conservation by wavelength pairs. Ultimately, for one set of wavelengths on-axis emission is observed exhibiting a strong overlap with the fundamental Gaussian mode, which can be efficiently coupled into single-mode optical fibers. Hence, proper design of the crystal length and coupling geometry is equivalent to frequency filtering [7] and may obviate the customary use of lossy band-pass filters. As derived in [7] the intrinsic bandwidth for a certain emission angle (spatially filtered solely by single-mode fiber coupling) scales as $\Delta\lambda \sim 1/L$. Furthermore, the authors of [7] conclude that, given optimal focusing conditions, the pair production rate also benefits from the crystal length and scales as $R \sim \sqrt{L}$. Hence, the spectral brightness B, a commonly used benchmark parameter for entangled photon sources, defined as photon pair number per unit of time [s], pump power [mW] and bandwidth [THz] relates to the crystal length as $B \sim L\sqrt{L}$.

In previous realizations [8], where two orthogonally oriented crystals were placed behind each other, rather short (<10 mm) crystals had to be used in order to preserve near-optimal focusing and fiber coupling conditions. Therefore, filtering with narrow-band interference filters was necessary in order to avoid undesirable effects of both chromatic dispersion (as described above) and polarization mode dispersion, which would lead to severe impairment of the quality of the polarization-entangled two-photon state upon transmission over long fibers. In the current work, the source incorporates only one periodically poled crystal allowing for a length of more than 30 mm, still keeping optimal focusing conditions.

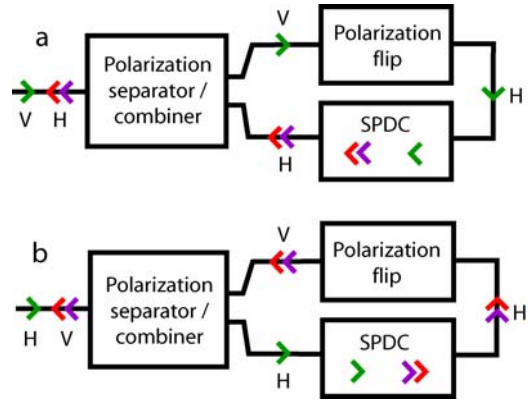

Fig. 1. Principle of the Sagnac interferometer. Depending on the polarization of the entering pump photon (green) the loop is passed in either clockwise (a) or counterclockwise (b) direction. The pump photons are transformed into signal (red) and idler (purple) by SPDC. The pump or the signal and idler photons are polarization flipped in cases a and b, respectively.

The Sagnac-type design (Fig. 1) has proved to be very successful in incorporating long crystals in SPDC sources producing entanglement. So far those sources were however limited to generation of (near) degenerate pairs of photons [9]. The reason stems from the difficulty to design a Sagnac loop capable of accepting a broad range of wavelengths as it is necessary for asymmetric SPDC. In particular the polarizing beamsplitter and the polarization flip in one arm have to work for all wavelengths present in the loop. In near degenerate setups only two wavelengths have to be matched, namely the pump wavelength and the signal/idler wavelength. It is therefore possible to use special dual beamsplitters and dual half-waveplates which are readily available from stock. Contrarily, in a three color Sagnac loop, needed for asymmetric SPDC, the choice of suitable broadband elements is limited and very costly.

A Sagnac-type source for entanglement at asymmetric wavelengths was recently reported [10], which worked around this problem by unfolding the Sagnac loop into two parts. Pump light (532 nm) and idler (1550 nm) traveled in the first loop, whereas the signal (810 nm) traveled in the second loop. This design loosened the constrictions on triple-wavelength components but lead to the loss of the inherent phase stability of the Sagnac interferometer. Since the two loops were spatially separated their relative phase was no longer constant and an active phase stabilization with nano-positioning devices had to be introduced.

We present here a new design of a Sagnac-based source for entanglement which works for a broad spectral range without the need of expensive optical components or any active stabilization. By leaving a planer structure and exploiting a three-dimensional beam path a simple solution for the three-color Sagnac loop was found as detailed in the next chapter. The careful design of the interferometer ensures self-compensation of phase drifts as well as exact balance of the two arms in terms of dispersion.

## 3. Source of polarization-entangled photons

To generate pairs of polarization-entangled photons by means of SPDC a Sagnac-type interferometer has been built. In this concept the two interfering modes occupy physically the same optical paths, but are passed in opposite directions. Therefore, in case of operation at a single wavelength only, exact phase compensation results. This is also valid for the free-space parts in our multi-wavelength scenario, due to the absence of material dispersion (neglecting the dispersion of air). However, inside optical elements mechanical drifts as well as temperature changes are self-compensated to first order approximation only, which shall be derived in the following:

The entangled two-photon state can be written as

$$|\phi\rangle = \frac{1}{\sqrt{2}}\left(|H\rangle|H\rangle + e^{i(\Phi_0 + \Delta\Phi)}|V\rangle|V\rangle\right) \quad (1)$$

where $\Phi_0$ is the global phase (e.g. zero for a $|\phi^+\rangle$ state) and $\Delta\Phi$ accounts for a (temperature induced) phase drift consisting of contributions of the three light fields involved:

$$\Delta\Phi = \Delta\varphi_s + \Delta\varphi_i - \Delta\varphi_p. \quad (2)$$

The negative sign arises from the counter-propagating direction of the pump component with respect to signal and idler and is the manifestation of the self-compensating effect. Each term in Eq. (2) can be written as

$$\Delta\varphi_x = \frac{2\pi L}{\lambda_x}\left(\frac{dn_x}{dT} + n_x\alpha\right)\Delta T, \quad (3)$$

where $x$ denotes signal, idler and pump, respectively, $\lambda$ is the wavelength, $n$ and $L$ are the refractive index and the length of the material, $\alpha$ is the thermal expansion coefficient, and $\Delta T$ is the temperature change. The material parameters shown in Tab. 1 are obtained from the temperature dependent Sellmeier equations [11]. Considering for instance the warming of a single optical element within the loop (e.g. 8 mm of BK7-glass) by $\Delta T = 1$ K we end up with a total phase shift of only $\Delta\Phi = 0.019\ \pi$. Hence, sufficient phase stability is ensured, even in our highly non-degenerate case. In comparison, a Mach-Zehnder-type interferometer providing no compensation would introduce a phase shift (given by Eq. (3) alone) of $\Delta\Phi = \Delta\varphi \approx 0.36\ \pi$ for the same conditions, and hence rendering the quantum state unusable for long term applications.

Table 1. Refractive index, its temperature dependence and thermal expansion coefficient of BK7-glass.

| $\lambda$ [nm] | $n$ | $dn/dT$ [K$^{-1}$] | $\alpha$ [K$^{-1}$] |
|---|---|---|---|
| 532 | 1.5195 | $1.55\times10^{-6}$ | |
| 810 | 1.5106 | $1.095\times10^{-6}$ | $7.1\times10^{-6}$ |
| 1550 | 1.5007 | $8.633\times10^{-7}$ | |

In our setup several optimizations concerning three-color operation, stability and space requirements have been put in place. All the optics are designed for broadband operation by means of total internal reflection (or grazing incidence in case of one silver mirror), thus the interferometer is potentially suitable for all wavelengths in the range of 400 nm - 2 μm. All optical elements are free from angular dispersion ensuring ease of alignment for both signal and idler beams. The two round-trip paths are balanced in terms of group delay for each of the three wavelengths, providing robust entanglement even for pulsed or low-coherence pump lasers. The polarization and wavelength dependent phase shift induced by the totally reflecting elements is static and can be included in the global phase ($\Phi_0$ in Eq. (1)). The complete interferometer base has been milled from a solid block of aluminum ensuring mechanical stability and having a footprint of 5 × 5 cm only. Stable operation (in terms of count rates as well as phase) has been observed over several hours and on a day-to-day basis.

The source is pumped by a 532 nm single frequency DPSS-laser (*Laser Quantum, Torus*). Special care has been taken to remove spurious wavelength components around 808 nm and 1064 nm by means of multiple reflections off dichroic mirrors.

Fig. 2 shows the schematic of the optical setup. A half-wave plate is used to adjust the ratio of pump power in the horizontal (H) and vertical (V) polarization direction. The relative phase between the H and V component is controlled with two birefringent quartz wedges, the longitudinal walk-off (group delay between coherent wave packets) of which is undone by an orthogonally oriented block of the same material and approximately the same length. A singlet lens (f = 166 mm) produces a beam waist of ~80 μm $1/e^2$-diameter precisely at the center of the ppKTP crystal. The two polarization modes are separated with a high extinction ratio ($1:10^5$) by a customized Glan-Thompson polarizer avoiding angular dispersion. Again its temporal walk-off is compensated for with a piece of calcite of equal length. The polarization of the vertically polarized beam is flipped by a cross-faced periscope (instead of a half-wave plate usually used), while the horizontally polarized beam is accordingly displaced in height by a parallel-faced periscope (insert in Fig. 1). Hence, the ppKTP crystal, lying on a lower level than the Glan-Thompson polarizer, is pumped from both sides with horizontally polarized light. The crystal is heated to 65°C in order to fulfill type-I phase matching conditions for 532 nm → 810 nm + 1550 nm. Since we work at highly non-degenerate wavelengths the signal and idler modes can be separated by means of their color. Hence, there is no need for two distinctive output modes directly from the source (as in a type-II down conversion scheme [1,9]) and the superior type-I nonlinear susceptibility of KTP can be utilized. The generated photon pairs continue their way through the Sagnac loop, one of them being polarization flipped in the cross-faced periscope. Subsequently they are recombined into one beam at the polarizer and reflected off the trichroic mirror. Finally the photons at the two wavelengths are separated by a dichroic mirror and launched into single-mode fibers with appropriate couplers. It should be stressed that both periscopes work purely on a geometrical effect and the polarizer is based on total internal reflection allowing those components to perform over a large spectral range. This fact allows for the Sagnac loop to be operated between 532 and 1550 nm.

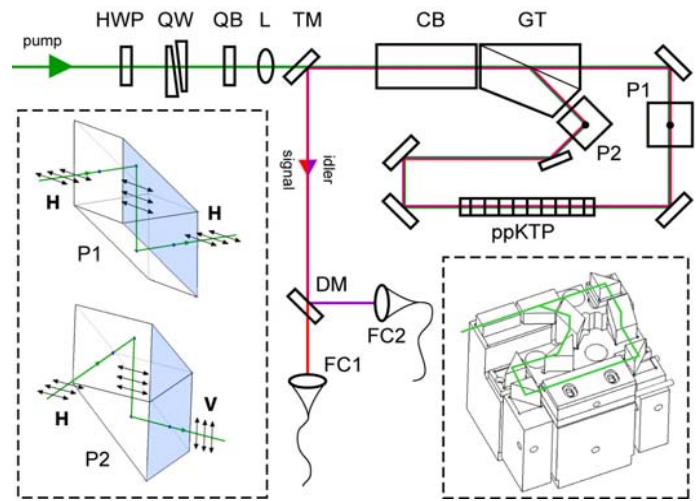

Fig. 2. Schematic of the source: half-wave plate (HWP), quartz wedges (QW), quartz block (QB), focusing lens (L), trichroic mirror (TM), calcite block (CB), Glan-Thompson polarizer (GT), parallel-faced periscope (P1), cross-faced periscope (P2), periodically poled Potassium Titanium Oxide Phosphate crystal (ppKTP), dichroic mirror (DM), fiber couplers (FC1 & FC2). Insert left: principle of operation of the two periscopes. Insert right: 3-dimensional view of the implemented interferometer.

## 4. Performance

The created photon pairs were characterized in detail (Fig. 3). The signal photons (810 nm) were detected using a free running home-built single-photon counting module incorporating a Si-APD with a quantum efficiency of ~30%. At an average pump power of P = 1 mW in each polarization (H and V) single-count rates of $R_s$ = 600000 s$^{-1}$ per mode have been measured with a dark count background of $R_d$ = 1000 s$^{-1}$. The idler photons (1550 nm) were detected with a commercial InGaAs based photon counting module (*id Quantique, id201*) set to a quantum efficiency of 15%. This detector was gated with a coincidence window of $\tau$ = 2.5 ns and triggered from the output of the Si-APD. Coincidence rates of $R_c$ = 9300 s$^{-1}$ were obtained for each of the two polarizations, giving a coincidence-to-singles ratio of 1.55%. The accidental coincidences were determined to be $R_a$ = 400 s$^{-1}$ by shifting the trigger off the coincidence peak.

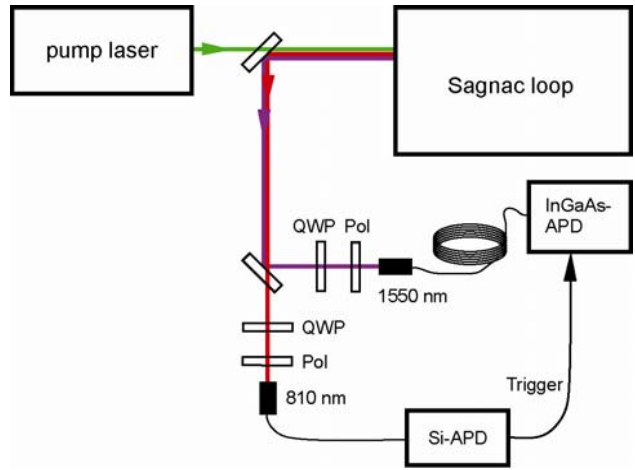

Fig. 3. Setup for efficiency measurement and quantum state tomography. The signal photon at 810 nm is detected by a Si-APD detector. A trigger signal is sent to an InGaAs-APD detector where the appropriately delayed 1550 nm idler photon can be detected. Quarter-wave plates and polarizers in both arms are used for the characterization of the polarization state.

The bandwidth of the signal photons was measured with a single-photon spectrometer yielding a value of $\Delta\lambda_s$ = 0.4 nm ($\Delta\nu_s$ = 180 GHz). This results in a brightness B = $R_c/(\eta_s \eta_i) \times 1/P \times 1/(\Delta\nu_s)$ = 1.13 × 10$^6$ s$^{-1}$ mW$^{-1}$ THz$^{-1}$ present in the optical fibers . Here, $\eta_s$ and $\eta_i$ are the respective detector efficiencies. The downconversion process forces the 1550 nm idler photons also to be of narrow bandwidth ($\Delta\nu_i = \Delta\nu_s$), corresponding to $\Delta\lambda_i$ = 1.5 nm. It is remarkable that for a transmission distance as large as L = 100 km over standard telecom fibers the chromatic dispersion (D = 18 ps/nm/km) causes a broadening of the coincidence peak to as small a value as $\tau$ = D × $\Delta\lambda$ × L = 2.7 ns. This additional timing jitter is close to the minimal detection window of a typical InGaAs-APD in contrast to previous work [8], where it was significantly wider. This underlines the perfect suitability of our idler photons for long-distance quantum communication.

In order to quantify the entanglement of the two-photon state a tomography measurement was carried out. Coincidences at 16 independent settings of polarization for the signal and idler, respectively were measured [12]. Thus, the density matrix of the quantum state was reconstructed (Fig. 4) and compared to the theoretical matrix of a $|\phi^+\rangle$ state. The analysis yields a fidelity F = 97.5 ± 0.15% (the overlap with the maximally entangled state). The error bar of this result was estimated by performing a 1000 run Monte Carlo simulation of the whole state tomography analysis, with Poissonian noise added to the count statistics in each run. When the background of accidental coincidences was subtracted the fidelity increased to F = 98.2%.

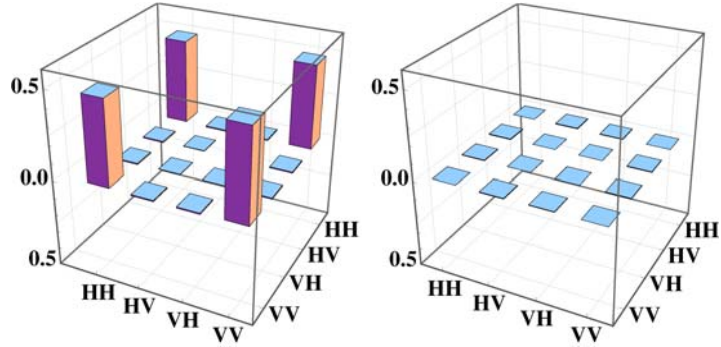

Fig. 4. Real and imaginary part of the reconstructed density matrix of the measured state. Accidental coincidences and dark counts have been subtracted.

## 4. Conclusion

We have demonstrated a polarization-entangled photon source at highly non-degenerate wavelengths using a modified Sagnac-type interferometer. To our knowledge this is the first time that such a design was employed for the production of polarization-entanglement at asymmetric wavelengths. Due to the Sagnac-type architecture, a passive self-compensation of phase is achieved. Our setup can make use of a single 30 mm long crystal, thereby reducing the bandwidth of the produced photons to 180 GHz. A tomographic measurement of the entangled state showed a fidelity of F = 97.5% at a detected coincidence rate of 9300 c/s per polarization mode. The source is very compact (5 × 5 cm) and robust making it a promising candidate for fiber based quantum information applications like quantum key distribution.

**Acknowledgments**

We would like to thank Sven Ramelow for discussions on the components of the polarization flip and Robert Prevedel for supplying invaluable support for the state tomography measurement. This work was supported by the Austrian Science Foundation FWF (TRP-L135 and SFB-1520).